\documentclass[aps,prb,twocolumn,groupedaddress]{revtex4-1}
\usepackage[dvips]{graphicx}
\usepackage{multirow}
\usepackage{bm}
\usepackage{verbatim}
\usepackage{lineno, blindtext, color}
\linenumbers

\begin{document}

\title{Dzyaloshinsky-Moriya driven helical-butterfly structure in Ba$_3$NbFe$_3$Si$_2$O$_{14}$}
\author{V. Scagnoli}
\email{valerios@ethz.ch}
\affiliation{Swiss Light Source, Paul Scherrer Institut, CH 5232 Villigen PSI, Switzerland}
\affiliation{ETH Z\"{u}rich, Institut f\"{u}r Quantenelektronik, W. Pauli Strasse 16, 8093 Z\"{u}rich, Switzerland}
\author{S. W. Huang}
\author{M. Garganourakis}
\author{R. A. de Souza}
\author{U. Staub}
\affiliation{Swiss Light Source, Paul Scherrer Institut, CH 5232 Villigen PSI, Switzerland}
\author{V. Simonet}
\author{P. Lejay}
\author{R. Ballou}
\affiliation{Institut N\'eel, CNRS and Universit\'e Joseph Fourier, BP166, 38042 Grenoble Cedex 9, France}

\date{\today}

\begin{abstract}
We have used soft x-ray magnetic diffraction at the Fe$^{3+}$ L$_{2,3}$ edges to examine to what extent 
the Dzyaloshinsky-Moriya interaction in Ba$_3$NbFe$_3$Si$_2$O$_{14}$ influences its low temperature 
magnetic structure. A modulated component of the moments along the $c$-axis 
is present, adding to the previously proposed  helical magnetic configuration of 
co-planar moments in the $a,b$-plane. This leads 
to a ''helical-butterfly'' structure and suggests that both the multi-axial in-plane and the uniform 
out-of-plane Dzyaloshinsky-Moriya vectors are relevant. A non zero orbital magnetic signal is also observed  
at the oxygen K edge, which reflects the surprisingly strong hybridization between iron 3$d$ and oxygen 2$p$ states, 
given the nominal spherical symmetry of the Fe$^{3+}$ half filled shell.

\end{abstract}



\maketitle

\section{Introduction}
The term chirality was first utilized in science by Lord Kelvin.
His original definition has evolved with time and we now speak about a 
chiral system if such a system exists in two distinct (enantiomeric) states that
are interconverted by space inversion, but not by time reversal combined with any proper spatial 
rotation.\cite{AvalosCR98}
Chirality permeates natural sciences from biochemistry to solid state physics.
The fact that living organisms use only the left enantiomers of amino acids is still not well understood.  
Chirality is also found in magnets.~\cite{Lovesey1996,SimonetEPJST213} 
An example is the left- or right- handedness associated with the helical order of magnetic moments.
In principle, the two states are degenerate, resulting in an equipopulation of chiral domains. 
However, competing interactions or external effects such as strain, can unbalance this ratio, favoring one particular state.
In particular, 
in non centrosymmetric crystals, characterized by the absence of parity symmetry, a single domain might be selected. 
Despite having 65 non centrosymmetric (including 22 chiral) space groups allowing chiral crystal structures, out of 230, only few single handed magnetic compounds were 
reported.~\cite{IshidaJPSJ54,MartyPRL101,GrigorievPRL102,JanoschekPRB81} 
Interest in such systems is two-fold. First, they can exhibit interesting physical properties such as magnetic 
Skyrmion lattices~\cite{MuhlbauerS323} or 
helimagnons.~\cite{JanoschekPRB81} The second is related to the discovery of magnetically induced multiferroics~\cite{KimuraN426} 
where researchers struggle to find materials with a stronger 
electrical polarization.~\cite{JohnsonPRL108} The latter is directly affected by the imbalance between chiral domains, which possess opposite 
electric polarizations. Therefore, materials 
showing a single chiral domain are promising candidates to host a significant macroscopic electrical polarization, 
which makes them an ideal model system to study. 
Ba$_3$NbFe$_3$Si$_2$O$_{14}$ gathered attention in this respect, exhibiting fully chiral magnetism~\cite{MartyPRL101}  and magnetoelectric coupling phenomena.~\cite{MartyPRB81,ZhouCM21,LeeCM22}

Ba$_3$NbFe$_3$Si$_2$O$_{14}$ crystallizes in a trigonal P321 space group ($a=b=8.539$, $c=5.241$, $\gamma=120^{\circ}$). 
It displays an antiferromagnetic order below $T_N$=27~$\!$K. The magnetic moments are localized on the Fe$^{3+}$ 
ions ($L \simeq 0$, $S=5/2$). These occupy the Wyckoff position (3f) (0.2496, 0, 0.5) 
with .2. site symmetry, forming triangular units in the $a$,$b$-planes. 
Elastic neutron scattering studies~\cite{MartyPRL101} suggest that the same triangular configuration of co-planar moments at $120^{\circ}$ from each 
other is stabilized within each triangle of an $a$,$b$-plane and that this arrangement is helically modulated from $a$,$b$-plane 
to $a$,$b$-plane along the $c$-axis according to the propagation vector $(0, 0, \tau)$ with $\tau$ close to 1/7 
(see Fig.~1 of Ref.~\onlinecite{MartyPRL101}). 
An extremely appealing discovery was that the single crystals are grown  enantiopure and that the low temperature magnetic structure 
is single domain, with a single chirality of the triangular magnetic arrangement on the triangles and a single chirality of the 
helical modulation of the magnetic moments, which was dubbed helicity.~\cite{MartyPRL101} 
It was suggested that the Dzyaloshinsky-Moriya~\cite{DzyaloshinskyJPCS4,MoriyaPRL4} exchange interaction might be responsible 
for selecting the ground state configuration~\cite{MartyPRL101} and for the opening of a small gap in the magnetic excitation 
spectrum.~\cite{LoirePRL106} 
Another inelastic neutron scattering study proposed the latter to arise from single ion anisotropy,~\cite{StockPRB83} 
but recent  spin resonance experiments support the first scenario indicating furthermore that not only the uniform 
component along the $c$-axis of the Dzyaloshinsky-Moriya vector but also its multiaxial component within the 
$a$,$b$-plane might be sizeable.~\cite{ZorkoPRL107} The latter could generate an additional component to the magnetic 
structure not necessarily detected by neutron scattering. 
To find evidence for such a magnetic motif we have used resonant 
x-ray diffraction at the Fe L edges. Our results show clear deviations from the magnetic structure previously proposed,  
confirming the existence of such a component.
\section{Experimental Details}
Powders of Ba$_3$NbFe$_3$Si$_2$O$_{14}$ were synthesized by solid state reaction from stoichiometric 
amounts	of Nb$_2$O$_3$, Fe$_2$O$_3$, SiO$_2$ oxides and BaCO$_3$ barium carbonate, at 1150$^{\circ}$ C in air within 
an alumina crucible. The reagents were carefully mixed and pressed at 1GPa to form compact cylinders before annealing. 
The phase purity was checked by x-ray powder diffraction. 
Single crystals were grown from the as-prepared polycrystalline cylinders by the floating-zone method in an image furnace.~\cite{BordetJPCM18} 
The single crystal used in the present investigation was extracted from the same batch as the one used in Ref.~\onlinecite{MartyPRL101} and 
has the same structural chirality $\epsilon_{T}$, to be precise $\epsilon_{T}=-1$. After polishing the surface perpendicular to the [001] direction it was annealed to improve the surface quality.

We have performed resonant x-ray diffraction experiments 
at the Fe L$_{2,3}$ edge. These energies correspond to a wavelength of approximately 
17~\AA~and are associated to an electric dipole resonance from the iron 2$p$ to 3$d$ levels. 
Experiments were performed
with the RESOXS chamber~\cite{StaubJSR15} at the X11MA beamline~\cite{FlechsigAIP1234} of the Swiss Light Source. 
The twin Apple undulators provide linear, horizontal $\pi$ and vertical $\sigma$, and circularly, right $R$ and left $L$, polarized x rays with a 
polarization rate close to 100\%. The polarization of the diffracted beam was not analyzed. 
The sample was attached to the cold finger of an He flow cryostat with a base
temperature of 10~$\!$K. Azimuthal scans were achieved by rotation of the single crystal, with an accuracy of approximately $\pm5^{\circ}$.
%
%
\section{Resonant x-ray Scattering}
The x-ray cross section for magnetic scattering is normally very small, though at synchrotron photon sources 
such weak signals are routinely measurable. 
However, when working close to an atomic absorption edge the magnetic scattering signals are significantly 
enhanced and are element sensitive. 
Resonant x-ray diffraction occurs when a photon excites a core electron to empty states, and is subsequently 
re-emitted when the electron and the core hole recombine.~\cite{HannonPRL61,HannonPRL62,HillACA52}
This process introduces anisotropic contributions to the x-ray susceptibility tensor,~\cite{TempletonACA38,DmitrienkoACA39,TempletonACA42}
the amplitude of which increases dramatically as the photon energy 
is tuned to an atomic absorption edge. In the presence of long-range magnetic 
order, or a spatially anisotropic electronic distribution, the interference of the anomalous 
scattering amplitudes may lead to Bragg peaks at positions forbidden by the
crystallographic space group. An example of such a resonant enhancement of the diffracted intensity as a function of energy occurring in the vicinity of the Fe L$_3$ edge  
 in Ba$_3$NbFe$_3$Si$_2$O$_{14}$ is given in Fig.~\ref{fig:Escanlangajuly}. 
X rays thus prove to be a valid 
alternative or complementary tool to neutron diffraction for the study 
of magnetic structures.~\cite{ScagnoliPRB73,WilkinsPRL103,JangPRL106,JohnsonPRL107,ScagnoliPRB86} 
Its superior resolution in reciprocal space can be advantageous, simplifying for instance the precise determination of 
incommensurate magnetic phases, which is relevant in cases where 
the incommensurability is very small.~\cite{AgrestiniPRB77}

To understand the content of the x-ray resonant magnetic cross section, it is customary
to use the expression first derived by Hannon and Trammell for an electric dipole (E1) event:~\cite{HannonPRL61,HannonPRL62,HillACA52}
\begin{equation}{\label{eq:hannon}}
F^{E1}_{\mbox{\boldmath $\epsilon$}^{\prime},\mbox{\boldmath $\epsilon$}} = 
                ({\mbox{\boldmath $\epsilon$}}^{\prime} \cdot {\mbox{\boldmath $\epsilon$}}) F^{(0)} 
               - i ({\mbox{\boldmath $\epsilon$}}^{\prime} \times {\mbox{\boldmath $\epsilon$}}) \cdot
	       \hat{\mathbf{z}}_n F^{(1)} 
	       + (\hat{\mbox{\boldmath $\epsilon$}}^{\prime} \cdot \hat{\mathbf{z}}_n)
	       (\hat{\mbox{\boldmath $\epsilon$}} \cdot \hat{\mathbf{z}}_n) F^{(2)} ,
\end{equation} 
where the first term contributes to the charge (Thompson) Bragg peak. 
The second and third terms correspond to magnetic diffraction.
$\hat{\mathbf{z}}_n$ is a unit vector in the direction of the magnetic moment of the $n$th ion in the unit cell and 
$\bm{\epsilon}$ ($\bm{\epsilon}^{\prime}$) describes the polarization state of the incoming (outgoing)
x rays. $F^{(i)}$ depend on atomic properties and determine the strength of the resonance.~\cite{HillACA52,Lovesey1996} 
In an antiferromagnet, the second term produces the first-harmonic magnetic satellites and the 
third term, which contains two powers of the magnetic moment, produces the second-harmonic magnetic satellites.
It shows how the intensity of the magnetic diffraction depends on the motif of the  magnetic moments and 
 on the orientation of the sample relative to the incident x-ray polarization state.
In particular, a non collinear magnetic motif is able to produce a different diffraction intensity 
depending on the helicity of the incident x rays, e.g. $I_{R} \neq I_{L}$, where $I_{R}$ is the intensity measured with 
incident right-handed circularly polarized photons and $I_{L}$ for left-handed ones. 
Rotating the sample about the diffraction wave vector might result in a smooth change of the diffracted intensity which 
helps to reconstruct the magnetic moment motif.

It is worth emphasizing that Eq.~(\ref{eq:hannon}) is an approximation for the resonant magnetic scattering 
cross section which, strictly speaking, is only valid for a cylindrical symmetrical environment of the resonant ion. 
When this approximation does not hold the diffracted intensities must be described as exemplified in 
Ref.~\onlinecite{CarraRMP66,Lovesey1996,LoveseyPR411,StojicPRB72,ScagnoliPRB79,HaverkortPRB82}. 
%
%
\section{Results}
\begin{figure}[t]
\includegraphics[scale=0.46,angle=0,bb = 20 181 560 600, clip]{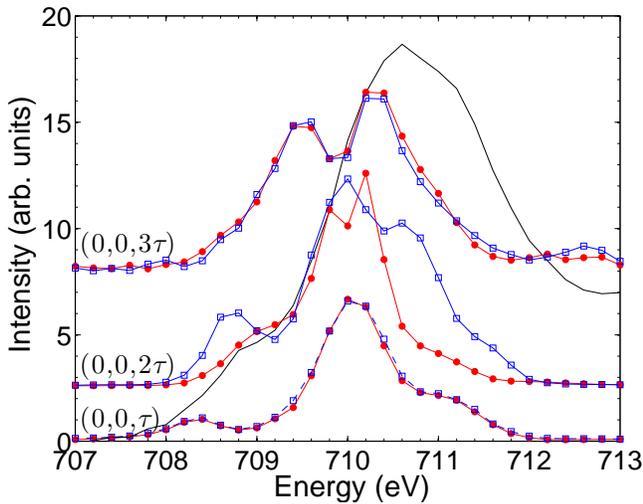}
\caption{\label{fig:Escanlangajuly} (Color online) Intensity versus energy of the three satellite reflections in the vicinity of the 
Fe L$_3$ edge. Spectra
collected with incident $\pi$ [(blue) square] and $\sigma$ polarizations [(red) filled circle] at 10~K. 
Spectra are scaled [(0, 0, $\tau$) and (0, 0, $3\tau$) were multiplied by 2.5 and 80 respectively]
 and shifted for clarity and lines are guides to the eye. 
The black continuous line represents the sample absorption spectra 
collected in fluorescent mode. }
\end{figure}

Once the sample is cooled below the N\'eel temperature $T_N$, superstructure peaks (0, 0, $n\tau$) of 
order $n$ up to three 
arise from magnetic ordering and magnetically induced lattice distortions.  
The observation of such reflections is remarkable as, given the 
magnetic motif suggested by neutron diffraction, they should be absent. 
They are of resonant nature and they disappear when the energy of the incident x ray is detuned from the iron L
edges (Fig.~\ref{fig:Escanlangajuly}). 
Non-resonant magnetic intensity could be zero or too small to be visible. 
Resonant x-ray diffraction is sensitive to the spin, orbital and charge degrees of
freedom.~\cite{MurakamiPRL80,HillACA52,StaubPRL88,StaubPRB71}
In order to assert their origin and refine the magnetic structure, we collected their energy, azimuthal and temperature dependence.
%
%
%
%
%
\begin{figure}[t]
\includegraphics[scale=0.46,angle=0,bb = 22   179   551   597, clip]{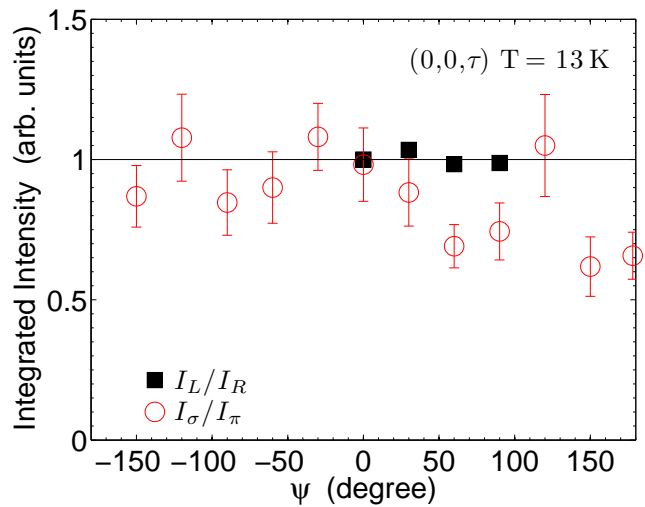}
\caption{\label{fig:azilanga} (Color online) Azimuthal angle dependence of the $I_{\sigma}/I_{\pi}$ (red) and 
$I_L/I_R$ (black) ratio for the (0, 0, $\tau$) magnetic reflection. 
The (black) line represents  the predictions of the model described in the text 
 ($\chi^2= 4.0$ for comparison with both dataset, 
$\chi^2= 1.5$ for the ratio $I_L/I_R$ alone). 
Measurements were performed in the vicinity of the Fe L$_3$ edge. The  
azimuthal angle equals zero when the [100] direction is in the scattering plane. }
\end{figure}

Fig.~\ref{fig:Escanlangajuly} shows the energy dependence of the three superstructural peaks collected for x rays with
polarization in the diffraction plane (so-called $\pi$ geometry) and perpendicular to it ($\sigma$ geometry). 
They measure the maximum intensity of the diffraction peak at different energies (i.e. energy scans at fixed momentum transfer).
The first harmonic peak ($n=1$) shows equal intensity ($I_{\pi}$ = $I_{\sigma}$) for both incident x-rays polarization as the energy of the incident x-rays
is swept across the iron L$_3$ edge.
The ratio $I_{\pi}$ over $I_{\sigma}$ is very close to one and has no significant modulation as the sample 
is rotated about the diffraction wave vector 
$(0, 0, \tau)$ (so-called azimuthal-angle rotation), as exemplified in Fig.~\ref{fig:azilanga}. 
Data are collected for a Bragg angle $\theta_{B}=14.1^{\circ}$ where a significant contribution from 
specular reflectivity is present. 
Such a contribution is different for $I_{\pi}$ and $I_{\sigma}$ and, combined with the weakness of the signal,
complicates the determination of the magnetic Bragg diffraction contribution. 
In this respect, the data gathered with incident circularly polarized photons ($I_R$ and $I_L$) provide a more 
reliable data set, as being  a complex combination of the linearly polarized light, they present the same 
background for $I_R$ and $I_L$. Indeed the ratio 
$I_L$ over $I_R$ is very close to one over the investigated range and sports smaller error bars.

\begin{figure}[t]
\includegraphics[scale=0.46,angle=0,bb = 25 179 546 595, clip]{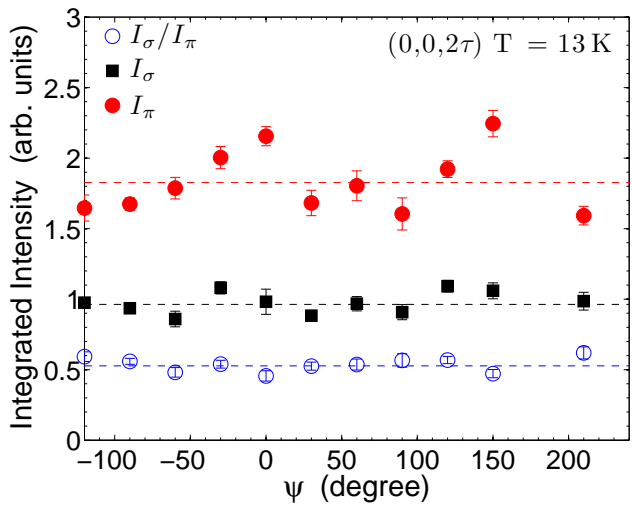}
\caption{\label{fig:azilanga2tau} (Color online) Azimuthal angle dependence of the (0, 0, 2$\tau$) superstructural reflection. 
The line represents a fit to the data with a constant ($\chi^2= 1.6$ for the ratio $I_{\sigma}/I_{\pi}$), as expected form the
model presented in the text. 
Measurements were performed in the vicinity of the Fe L$_3$ edge (E=709.4 eV). The  
azimuthal angle equals zero when the [100] direction is in the scattering plane. }
\end{figure}

The second harmonic (0, 0, 2$\tau$) energy dependence  has $I_{\pi}$ $\neq$ $I_{\sigma}$. Being associated with small
lattice or electron density deformations induced by the magnetic ordering, it is expected to 
exhibit a $I_{\sigma}$/$I_{\pi}$ ratio different from one. 
We do not observe any intensity far from the absorption edge. It indicates that the signal originates from the 
asymmetry of the electron density that appears below the magnetic ordering temperature, possibly triggered by 
the antiferromagnetic ordering.  
We have also collected its azimuthal angle dependence (Fig.~\ref{fig:azilanga2tau}). In analogy with the first harmonic peak 
it
shows no modulation, with $I_{\sigma}$ and $I_{\pi}$ constant within the error bars. 
Such results are supported also by the azimuthal variation of the ratio $I_{\sigma}$ over $I_{\pi}$ which 
displays smaller error bars due to the elimination of possible systematic errors, which equally affect both intensities, 
such as misalignments and changes in the sample illuminated area during the azimuthal scan.\\
Finally we discuss the third harmonic reflection (0, 0, 3$\tau$). 
Its energy dependence is quite peculiar. Being $I_{\pi}$ equal to $I_{\sigma}$
suggests the peak to be of magnetic origin, as in the case of (0, 0, $\tau$) reflection.
However, the spectral shape differs strongly from the one of the fundamental harmonic. It presents two principal features close in energy
rather than a single peak with two shoulders as in the case of the (0, 0, $\tau$).
As the iron site symmetry (.2.) does not forbid mixed events (e.g. electric dipole-quadrupole) one possible explanation
can be a small contribution coming from the electric quadrupole or electric dipole-quadrupole event, 
though such contributions are usually expected to be negligible. 
Note that the odd reflection intensities are indeed very small compared to other magnetic ordering signal found in
oxides.~\cite{WilkinsPRL91,StaubPRB71-214421,ScagnoliPRB73,GarciaFPRB78,StaubPRB80,deSouzaPRB84014409,JangPRL106} 
Effect of absorption correction can be discarded
as they would influence more significantly the (0, 0, $\tau$) reflection. At lower angles the penetration length is
reduced as the x rays have to travel longer into the sample before being diffracted into the detector. 
It was unfortunately not possible to collect its azimuthal angle dependence due to the weakness of the signal.
%

\begin{figure}[t]
\includegraphics[scale=0.46,angle=0,bb = 20 181 560 600, clip]{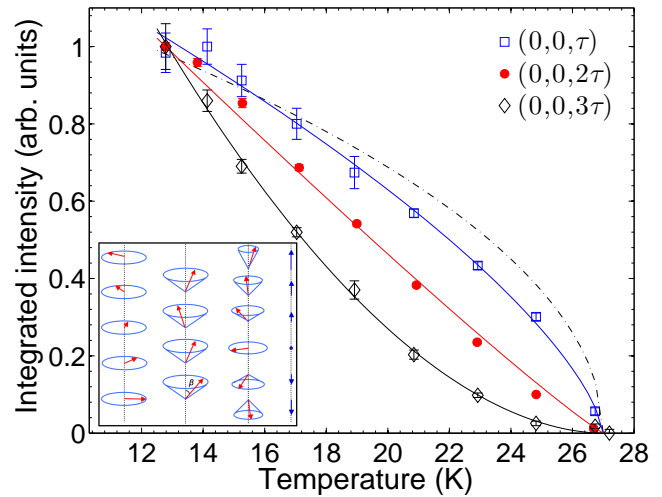}
\caption{\label{fig:Tdepend} Normalized integrated intensity vs temperature of the three satellite reflections. The solid lines show 
the best fit to power-law behavior $I_{n\tau} \propto (T_{N}-T)^{2\beta_{n}}$. The dashed line is the expected mean-field theory
dependence. The (0, 0, 2$\tau$) satellite is 7 times more
intense than the (0, 0, $\tau$). The same ratio holds between the (0, 0, $\tau$) and the (0, 0, 3$\tau$) satellites. 
The inset (left to right) shows different types of magnetic ordering: a simple spiral, a ferromagnetic (conical) spiral, a complex
spiral (or butterfly) and a static longitudinal wave. Data was measured with $\pi$ incident photon energy of 710 eV.}
\end{figure}

The temperature dependence of the satellite reflections (Fig.~\ref{fig:Tdepend}) shows strong resemblance to the one observed in rare-earth 
metals.~\cite{HelgesenPRB52} Pursuing the parallel with the rare-earth metals we would expect that the first harmonic 
arises from magnetic diffraction at the dipole resonance. The second harmonic corresponds to charge or orbital 
diffraction arising from
lattice or electron density modulations. The third-order harmonic might be a magnetic harmonic of the first or might originate from an electric
quadrupole resonance, although such a contribution is expected to be orders of magnitudes weaker. Our estimate of the
critical exponent $\beta$ found that it  is not consistent with mean-field theory.
A fit to power-law behavior $I_{n\tau} \propto (T_{N}-T)^{2\beta_{n}}$ gave an estimate  for the critical exponents. 
They are respectively $\beta_{1}=$0.34$\pm0.04$,  $\beta_{2}=$0.54$\pm0.05$, $\beta_{3}=$0.93$\pm0.08$.

Given the long modulation period of the magnetic structure it was possible to extend our 
investigation also to the oxygen K edge, which corresponds to an electric dipolar transition from the 1$s$ to the 
$2p$ level. 
\begin{figure}[t] 
\includegraphics[scale=0.44,angle=0,bb = 25 179 563 595, clip]{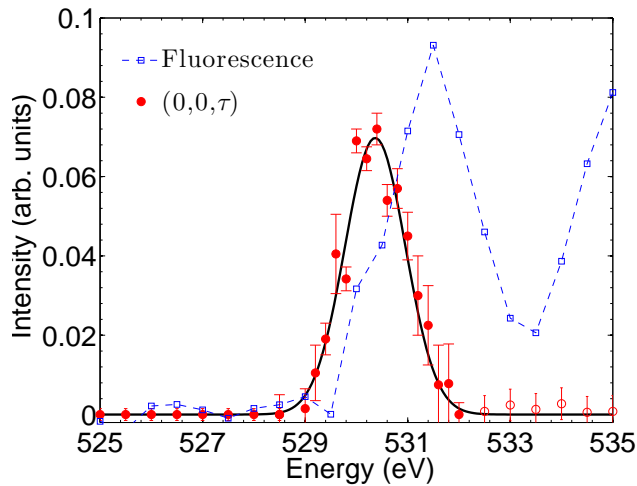}
\caption{\label{fig:langaOxy} (Color online) Intensity [(red) circle] vs energy of the (0, 0, $\tau$) satellite reflection at the
oxygen K edge collected at 10~K with $\pi$ incident x rays. The fluorescence spectra [(blue) open square] obtained in
the vicinity is also shown. Full (red) circle results from a fit of the integrated intensity of a reciprocal lattice
scan along the c* reciprocal lattice direction. Open (red) circle are a result of an energy scan with fix momentum
transfer. The (black) continuous line is a Gaussian fit of the oxygen resonance with a FWHM = 1.4$\pm$0.1 eV. }
\end{figure}
Upon cooling below $T_{N}$ a signal is observed at this energy. 
Figure~\ref{fig:langaOxy} shows its resonant nature. Observation of a resonant signal on an anion is not 
unusual.~\cite{MannixPRL86,BealePRL105,deSouzaPRB84} 
A resonant signal can arise, given a non zero overlap between the initial and the final
state, whereas a difference exists in the up/down spin dipolar overlap integrals. The difference can 
be induced by polarization of the orbitals.~\cite{vanVeenendaalPRL78} Such an asymmetry can arise also in case 
of a difference in the lifetime of the up/down spin channels.  
Recently Beale $et$ $al.$~\cite{BealePRL105} observed a resonant signal at the oxygen K edge in TbMn$_2$O$_5$, 
which they interpreted as a signature of an antiferromagnetically ordered spin polarization on the oxygen site. 
Such an observation is quite remarkable and we share their opinion that the study of oxygen spin polarization 
may lead to new insight in the understanding of the magnetoelectric coupling mechanism. 
As a matter of fact, an antiferromagnetic order at the oxygen site is consistent with neutron diffraction 
experiments that have already suggested a spin polarization of the oxygen by finding a value of 4 $\mu_B$ 
instead of the expected  5 $\mu_B$ for the spherical Fe$^{3+}$ half filled ion magnetic moment.~\cite{MartyPRL101,MartyPRB81} 
In our case the signal at the oxygen K edge is 90 times weaker than the corresponding one observed at the 
iron L$_3$ edge. Note that at the K edge the signal originates solely from the orbital magnetic moment component. 
No intensity was observed at the (0, 0, 2$\tau$) and (0, 0, 3$\tau$) satellites at the oxygen K edge. 
%
%
%
%
\section{Discussion}
Insights into the results can be obtained from group representation analysis,~\cite{BallouOuladdiaf2006}
provided that a single irreducible representation is selected at the magnetic ordering. The analysis is simplified by the 
fact that the space group P321 associated with the paramagnetic phase is symmorphic. It is, 
to be precise, a semi-direct product of the abelian translation group associated with a hexagonal 
lattice and the dihedral point group 32, which consists of the identity $1$, the anti-clockwise rotation $3^+$ and 
the clockwise rotation $3^- = (3^+)^2$ about the ternary $c$-axis and the dyads ($\pi$-rotations) about 
the three binary axes at $120^\circ$ to each other within the $a$,$b$-plane. 
A vector along the reciprocal $c^\star$-axis is reversed under the dyads and is invariant otherwise. 
It follows that the star of the magnetic propagation vector consists of the two 
vectors $\vec{\tau}_\pm = (0,0,\pm\tau)$ each being associated with the little space group P3, which is a semi-direct product of the translation group of the paramagnetic phase and the abelian 
cyclic point group 3.
An abelian group $G$ of $n_G$ elements has $n_G$ conjugacy classes 
(each being reduced to a singleton owing to the commutativity), which implies that it has 
$n_G$ irreducible representations $\Gamma_i ~(i = 1,...\,, n_G)$. It follows that these are necessarily 
all of dimension $d_i = 1$, to comply with the identity $\sum_{i=1}^{n_G} d_i^2 = n_G$. 
Each $\Gamma_i$ coincides then with its character $\chi_i$. The value of $\chi_i$ on any group 
element $g$ is an $n_G$-th root e$^{i 2\pi p/n_G} ~(p=1,...\,, n_G)$ of 1, because the order 
of $g$ always divide $n_G$. The character table is then built by making use of the 
orthogonality theorems. The basis vector of the invariant subspace of each $\Gamma_i$ is also easily 
deduced by applying the projection operator $\mathcal{P}_i = \frac{d_i}{n_G}\sum_{g \in G} \chi_i(g)^\star g$ on trial vectors.
Table~\ref{charactertableof3} summarizes such results for the cyclic group 3.
\begin{table}[b]
  \centering 
\begin{ruledtabular}
\begin{tabular}{c|ccc|c} &  & Characters &  &  Basis Vectors \\  & $1$ & $3^+$ & $3^-$ & \\ \hline $\Gamma_{1}$ & 1 & 1 & 1 & $\sum_{\nu=1}^3\vec{v}_\nu(\theta,\phi + (\nu-1)\frac{2\pi}{3})$ \\$\Gamma_{2}$ & 1 & $e^{i \frac{2\pi}{3}}$ & $e^{i \frac{4\pi}{3}}$ &  $\sum_{\nu=1}^3  e^{-i (\nu-1) \frac{2\pi}{3}} \vec{v}_\nu(\theta,\phi + (\nu-1)\frac{2\pi}{3})$ \\$\Gamma_{3}$ & 1 & $e^{i \frac{4\pi }{3}}$ & $e^{i \frac{2\pi}{3}}$ & $\sum_{\nu=1}^3  e^{-i (\nu-1) \frac{4\pi}{3}} \vec{v}_\nu(\theta,\phi + (\nu-1)\frac{2\pi}{3})$\end{tabular}
\end{ruledtabular}
  \caption{Irreducible representations of the cyclic point group 3, little co-group of the propagation vectors  
  $\vec{\tau}_\pm = (0,0,\pm\tau)$ in the space group P321, and associated invariant basis vectors. 
  $\vec{v}_\nu(\theta,\phi)$ symbolizes a vector associated to a Bravais lattice $\mathcal{L}_\nu$ at 
  an angle $\theta$ from the $c$-axis and the projection of which in the perpendicular plane is at an angle $\phi$ from the $a$-axis.}
  \label{charactertableof3}
\end{table}

The choice of a propagation vector amounts to select an irreducible representation of the translation group and 
determines a dephasing of moments within each Bravais lattice. Information on the phase relations between moments 
of distinct Bravais lattices can be extracted only from the irreducible representations of the little co-group. 
Three Bravais lattices $\mathcal{L}_\nu$ ($\nu$ = 1, 2, 3) are associated with the positions $(0.2496,~0,~0.5)$, 
$(0,~0.2496,~0.5)$ and $(-0.2496,-0.2496,~0.5)$ of the Fe$^{3+}$ ions on the 3f site. 
Under the symmetry operation 3$^+$ a moment of $\mathcal{L}_1$ (resp. $\mathcal{L}_2$, $\mathcal{L}_3$)  is rotated by an 
angle of 120$^\circ$ about the $c$-axis and is transported into $\mathcal{L}_2$ (resp. $\mathcal{L}_3$, $\mathcal{L}_1$) 
whereas under the symmetry 3$^-$ it is rotated by an angle of 240$^\circ$ about the $c$-axis and is transported into 
$\mathcal{L}_3$ (resp. $\mathcal{L}_1$, $\mathcal{L}_2$). 
This defines a representation $\Gamma$ of the cyclic group 3 of dimension 9 whose character $\chi$ takes the 
values $\chi(1) = 9$, $\chi(3^+) = 0$ and $\chi(3^-) = 0$ on the group elements. 
$\Gamma$ reduces into irreducible components as : $\Gamma = 3\Gamma_1 \oplus 3\Gamma_2 \oplus3\Gamma_3$. 
A magnetic structure can be most generally regarded as composed of several sine-wave amplitude modulations of moments: 
$\frac{1}{2}(\vec{v}_\nu(\theta_\nu, \phi_\nu)e^{-i \xi_\nu}e^{-i \vec{\tau}_\pm \cdot \vec{r}_{\nu n}} + c.c.)$,
where $\vec{r}_{\nu n} = \vec{r}_{\nu} + \vec{R}_n$ defines the position of the moment 
of $\mathcal{L}_\nu$ in the n-th unit cell, $\xi_\nu$ stands for an initial phase and
c.c. means to take the complex conjugate. 
The reduction of $\Gamma$ then suggests that, whatever the selected irreducible representation $\Gamma_i$, three independent directions of the moments are allowed by symmetry and can be combined, for instance along two orthogonal unit vectors in the $a$,$b$-plane, $\hat{x}_\nu = (\pi/2,  \phi_\nu)$ at an angle $\phi_\nu$ from the $a$-axis and $\hat{y}_\nu = (\pi/2,  \phi_\nu+\pi/2)$  at an angle $\phi_\nu+\pi/2$ from the $a$-axis, and along the unit vector $\hat{z}_\nu = (0, 0)$ of the $c$-axis, with possibly vectors $\vec{v}_\nu(\theta_\nu, \phi_\nu)$ of different lengths.

It was shown,~\cite{MartyPRL101} from collected neutron diffraction intensities, that a helicoidal modulation is stabilized within each $\mathcal{L}_\nu$, associated with a combination of the form 
$\vec{v}_\nu(\pi/2, \phi_\nu)e^{-i \xi_\nu} + \sigma \epsilon_H \vec{v}_\nu(\pi/2, \phi_\nu + \pi/2)e^{-i (\xi_\nu - \pi/2)} = m_{a,b} (\hat{x}_\nu + i \sigma \epsilon_H \hat{y}_\nu) e^{-i \xi_\nu}$ with $\sigma = +1$ for $\vec{\tau}_+$ and $\sigma = -1$ for $\vec{\tau}_-$. 
It is implicitly assumed that the vectors $\vec{v}_\nu(\pi/2, \phi_\nu) = m_{a,b}\hat{x}_\nu$ and $\vec{v}_\nu(\pi/2, \phi_\nu+\pi/2) = m_{a,b}\hat{y}_\nu$ have the same 
length $m_{a,b}$, which leads to a circular helix. An elliptic helix would have been obtained otherwise, 
which  {\it a priori} cannot be excluded.
$\epsilon_H = \pm 1$ defines the magnetic helicity, that is to say the sense of the rotation of the moments in the helix as one moves along the propagation vector: $\vec{m}(\vec{r}_{\nu n}) \times \vec{m}(\vec{r}_{\nu n} + \vec{c}) = \epsilon_H m_{a,b}^2 \sin(2\pi\tau) (\vec{c}/|~\vec{c}~|)$ whatever 
the chosen description between $\vec{\tau}_+$ and $\vec{\tau}_-$. 
If we impose $\phi_{\nu=2,3} - \phi_1$ according to Table~\ref{charactertableof3} then we must have $\xi_1=\xi_2=\xi_3$, which can be set to $0$, together with $\phi_1$, without loss of generality. 
Table~\ref{charactertableof3} illustrates that a triangular configuration of the moments on each triangle is associated with $\Gamma_1$ with a magnetic triangular chirality $+1$, that is to say with an anti-clockwise sense of the rotation of the moments as one moves anti-clockwise on a triangle. 
A triangular configuration of the moments on each triangle with the opposite magnetic triangular chirality $-1$, that is to say with a clockwise sense of the rotation of the moments as one moves anti-clockwise on a triangle, emerges from $\Gamma_2$ (resp. $\Gamma_3$) when $\epsilon_H = +1$ (resp. $\epsilon_H = -1$), in which case $\Gamma_3$ (resp. $\Gamma_2$) describes a ferro-collinear configuration of the moments on each triangle.
Intensity asymmetry of the pairs $\vec{K}\pm\vec{\tau}$ of magnetic satellites about reciprocal nodes $\vec{K}$ indicated that for, a left-handed structural chirality $\epsilon_T = -1$, if $\epsilon_H = -1$ then $\Gamma_1$ is selected and if $\epsilon_H = +1$ then $\Gamma_2$ is selected. 
This interdependence of the dephasing of moments within and between the Bravais lattices $\mathcal{L}_\nu$ was explained as arising from the twist in the exchange paths connecting the moments of consecutive $a$,$b$-planes, which depends on the structural chirality $\epsilon_T$ and imposes the magnetic triangular chirality $\epsilon_T \epsilon_H$. 
X-ray anomalous scattering confirmed that the structural chirality of the investigated crystal is $\epsilon_T = -1$. 
Neutron spherical polarimetry finally demonstrated that only the magnetic helicity $\epsilon_H = -1$, and therefore only the $(\epsilon_H, \epsilon_T\epsilon_H) = (-1, +1)$ magnetic helicity-triangular chirality pair, is selected, which was ascribed to the uniform Dzyaloshinsky-Moriya interactions with the Dzyaloshinsky-Moriya vectors all along the $c$-axis.
This model~\cite{MartyPRL101} was later confirmed by polarized neutron inelastic scattering with polarization analysis, which allowed probing both the symmetric and antisymmetric nature of the dynamical correlations associated with the magnon excitations emerging from the magnetic order.~\cite{LoirePRL106}

A crucial point of the reported model of the circular helices with moments within the $a$,$b$-plane is that the dephasing of the moments associated with the triangular configuration of moments on each triangle leads to zero magnetic structure factors at the scattering vectors $(0,0,\pm\tau)$. One however may recall  that the neutrons detect only the components of the moments perpendicular to the scattering vectors. An additional sine-wave amplitude modulated component along the $c$-axis of the moments is therefore not to be excluded, in which case we would rather have the combination $\vec{v}_\nu(\pi/2, \phi_\nu)e^{-i \xi_\nu} + \sigma \epsilon_H \vec{v}_\nu(\pi/2, \phi_\nu+\pi/2)e^{-i (\xi_\nu - \pi/2)} + \vec{v}_\nu(0, 0)e^{-i \xi_\nu^\prime}= m_{a,b} (\hat{x}_\nu + i \sigma \epsilon_H \hat{y}_\nu)e^{-i \xi_\nu} + m_c \hat{z}_\nu e^{-i \xi_\nu^\prime}$. The length $m_c$ of the vector $\vec{v}_\nu(0, 0)$ should however be small enough so that the neutron intensities to which it should give rise at the other scattering vectors, $(h,k, \ell\pm\tau)$ with $h \neq 0$ or $k \neq 0$, are drowned beneath the statistical uncertainties of the neutron intensities associated with the main helical modulation component. 
Table~\ref{charactertableof3} actually illustrates that this $c$-component of the moments would lead to a zero magnetic structure factor for the scattering vectors $(0,0,\pm\tau)$, and therefore would no longer be detected by resonant x-ray scattering, if the stabilized irreducible representation is either $\Gamma_2$ or $\Gamma_3$. A non-zero magnetic structure factor vectorially oriented along the $c$-axis is computed only in the case of the irreducible representation $\Gamma_1$: $F_m^{\Gamma_{1}}= (0,0,f_z)$. The magnetic intensity $I_{\epsilon^{\prime}\epsilon}=F^{\phantom{1}}_{\epsilon^{\prime}\epsilon} F^{*}_{\epsilon^{\prime}\epsilon}$ 
($^{*}$ stands for complex conjugation) in the different diffraction channels~\cite{vectordrop}  
 ($\epsilon = \sigma,\pi$ and $\epsilon^{\prime} = \sigma^{\prime},\pi^{\prime}$), associated with this amplitude modulated $c$-component, can be calculated with the help of Eq.~(\ref{eq:hannon}) leading to 
\begin{eqnarray}{\label{eq:fgamma1}}
I_{\sigma^{\prime}\sigma}  &=&  I_{\pi^{\prime}\pi } = 0 \,, \\ \nonumber
I_{\pi^{\prime}\sigma} &=& I_{\sigma^{\prime}\pi } \propto  \sin^{2} \theta_B \,.
\end{eqnarray}
where $\theta_B$ is the Bragg angle.
Noteworthy is the absence of any azimuthal dependence. We therefore expect no modulation of the intensity as we rotate the sample about the scattering wave vector. Moreover, we expect 
$I_{\sigma}=(I_{\sigma^{\prime}\sigma}+I_{\pi^{\prime}\sigma})=I_{\pi}=(I_{\sigma^{\prime}\pi }+I_{\pi^{\prime}\pi })$ and $I_R = I_L$. 
The latter equality can be derived from Eq.~(A1) in Ref.~\onlinecite{FernandezRPRB77} which states 
$I_R-I_L=$Im$\{F^{*}_{\sigma^{\prime}\pi} F_{\sigma^{\prime}\sigma} + F^{*}_{\pi^{\prime}\pi } F_{\pi^{\prime}\sigma} \}$.

Another deviation of the magnetic structure might arise from a slight ellipticity of the helices, but according to Table~\ref{charactertableof3} this would remain invisible in the case of the irreducible representation $\Gamma_1$. A finite magnetic structure factor, either $F_m^{\Gamma_{2}}= (f_x,f_y,0)$ or $F_m^{\Gamma_{3}}= (f_x^{\prime},f_y^{\prime},0)$, for the scattering vectors $(0,0,\pm\tau)$  would be obtained only if either the $\Gamma_2$ irreducible representation or the $\Gamma_3$ irreducible representation were to be stabilized as the main helical modulation component of the magnetic structure, but this is ruled out from the neutron diffraction data. 

A mixing of the irreducible representation $\Gamma_1$ with the irreducible representation $\Gamma_2$ (or $\Gamma_3$) finally is {\it a priori} not to be excluded, though this would imply that the magnetic transition is necessarily first order. 
Nevertheless, the additional magnetic component should be extremely tiny to escape standard powder neutron detection, since it should lie in the $a$,$b$-plane to produce a non zero magnetic structure factor.
In the case of the ferro-collinear configuration in the  $a$,$b$-plane, associated with irreducible representation $\Gamma_2$ for $\epsilon_H = -1$, which gives rise to a magnetic structure factor of the form $F_m^{\Gamma_{2}}= (f_x,f_y,0)$, one calculates with the help of Eq.~(\ref{eq:hannon}) the intensities:
\begin{eqnarray}{\label{eq:fgamma3}}
I_{\sigma^{\prime}\sigma}  &=&0 \,, \\ \nonumber
I_{\pi^{\prime}\sigma}  &=&I_{\sigma^{\prime}\pi }  =    k_1  \cos^2 \theta_B \,, \\\nonumber 
I_{\pi^{\prime}\pi } &=&  k_2  \sin^{2}(2\,\theta_B)  \, ,
\end{eqnarray}
where the constants $k_i$ depend on the amplitude of the component of the moments associated with the irreducible representation $\Gamma_2$ and their orientation in the $a$,$b$-plane with respect to the moments associated with the main irreducible representation $\Gamma_1$. 
Even in this case there is no azimuthal angle dependence, but we find $I_{\sigma} < I_{\pi}$ and $I_R \neq I_L$.
Including both $\Gamma_1$ and $\Gamma_3$ contributions will lead to an azimuthal angle dependence in the rotated 
channels and again $I_{\sigma} < I_{\pi}$ and $I_R \neq I_L$.

We are now in the position to compare the x-ray experimental data with the prediction from representation theory.
Fig.~\ref{fig:azilanga} shows that the ratio $I_R$ over $I_L$ is constant as a function of the azimuthal angle and equals one. 
Also the ratio $I_{\sigma}$ over $I_{\pi}$ is roughly constant within the error bars and is very close to one. 
It is thus clear that no mixing of irreducible representations is detected and that the magnetic structure abides by only the 
irreducible representation $\Gamma_{1}$ but involves components of the moments along the three orthogonal direction in space. 
As a whole it consists of moments in a triangular arrangement on each triangle in the $a$,$b$-plane helically modulated along 
the $c-$axis and exhibiting small up and down oscillations along the $c$-axis in phase with each other and with the 
same period as the helical modulation. Such a motif is reminiscent of the beatings of butterfly wings (although these wings 
here are three in number and not four), that lead us to dub it as ''helical-butterfly''. The existence of the butterfly component 
is consistent with the Dzyaloshinsky-Moriya interactions. Owing to the presence of the three 2-fold axes at $120^\circ$ of 
each other in the $a$,$b$-plane, each being perpendicular to one of the three sides of every triangle of moments, the 
Dzyaloshinsky-Moriya vector associated with each pair of moments must by symmetry lie within the plane containing the 
link connecting the two moments.~\cite{MoriyaPR120} The Dzyaloshinsky-Moriya vector field may therefore have a uniform component 
along the $c$-axis and a multi-axial component along the side of each triangle. It is this last component that 
gives rise to the butterfly component. It has been suggested that its contribution might be significant~\cite{JensenPRB84} if 
not dominating.~\cite{ZorkoPRL107}

Let us now analyze the azimuthal-angle dependence of the (0, 0, 2$\tau$) reflection. 
According to the $\Gamma_{1}$ magnetic structure factor $F_m^{\Gamma_{1}}= (0,0,f_z)$ and the formalism 
to calculate magnetic diffraction intensity in Ref.~\onlinecite{HillACA52} we should observe intensity only in the 
unrotated ${\pi^{\prime}\pi }$ scattering channel which is at odds with the data shown in Fig.~\ref{fig:azilanga2tau}. 
To reconcile the observations with theoretical prediction we must adopt a more sophisticated model which does not 
rely on the fact that the  resonant ion environment is cylindrically symmetrical. We need a tensorial structure factor 
$\Psi^K_Q$ where the positive integer $K$
is the rank of the tensor, and the projection $Q$ can take the $(2K+1)$
integer values which satisfy $-K \leq Q \leq K$. For a dipole
transition, tensors up to rank 2 contribute ($K \leq 2$). $K=0$
reflects charge contribution, $K=1$ time-odd dipole, and $K=2$ time-even
quadrupole. For our superstructural reflection we are interested in the quadrupolar contribution and 
given the presence of the 3-fold axis parallel to the $c$-axis we have  
$\Psi^K_Q (0,0,2\tau) = (-1)^{2\tau} \langle T^K_Q \rangle [ 1+2 \cos(Q \alpha)]$ 
which is non-zero only for $Q=0$. 
$\langle T^K_Q \rangle$ is an atomic tensor that describes the contribution of each atom to the structure factor. 
Making use of the formula in appendix C of
Ref.~\onlinecite{ScagnoliPRB79} we obtain the following results for the structure factor in the 
different polarization channels: 
\begin{eqnarray}{\label{eq:f2tau}}
F_{\sigma^{\prime}\sigma}  &=& - \frac{1}{\sqrt{6}} \Psi^2_0  \,, \\ \nonumber
F_{\pi^{\prime}\sigma}  &=&F_{\sigma^{\prime}\pi }  =  0  \,, \\\nonumber 
F_{\pi^{\prime}\pi } &\propto&  \frac{1}{\sqrt{6}}(1+\cos^{2} \theta_B) \, \Psi^2_0 \,, \\\nonumber
F_{\sigma} / F_{\pi } &=& -1/(1+\cos^{2} \theta_B) \,.
\end{eqnarray}
A derivation of such relations is presented in the Appendix. 
Such a model suggests no azimuthal dependence in all the diffraction channels and  
a ratio $I_{\sigma} / I_{\pi }=0.6$ in relative agreement with the azimuthal dependence shown 
in Fig.~\ref{fig:azilanga2tau} with a $\chi^2=6.1$. Agreement is improved ($\chi^2=2.2$) by letting the ratio 
value vary as a free parameter, with the experimental value of 0.54$\pm 0.02$, still reasonably close to the 
one derived by Eq.(\ref{eq:f2tau}). 
However, such a ratio, as exemplified in Fig.~\ref{fig:Escanlangajuly}, is 
not constant as a function of energy. These deviations might arise from a small symmetry break resulting in 
a loss of the 3-fold axis which would cause extra terms to appear in the structure factor.
Experimental uncertainties are however too big to extract more quantitative conclusions on the presence of such 
contributions.
%
%
%
%
%
%
%
\section{Conclusion}
We have studied the magnetic structure of the intriguing compound Ba$_3$NbFe$_3$Si$_2$O$_{14}$ 
with resonant x-ray diffraction at the Fe L edges and O K edge. 
These experiments give new insight into the details of the magnetic structure recently  determined by neutron diffraction. 
Our experiments have found an extra sinusoidal modulation of the Fe magnetic moments along the crystallographic
$c$-axis, concomitant with the helical order in the $a$,$b$-plane, generating an 
helical-butterfly magnetic structure. 
Such sinusoidal modulation arises from the Dzyaloshinsky-Moriya interaction as suggested by symmetry consideration and
recent linear spin-wave theory calculations.~\cite{JensenPRB84} 
The orbital magnetic signal observed  
at the oxygen K edge reflects the strong hybridization between iron 3$d$ and oxygen 2$p$ states. 
Finally, the energy dependence of $I_{\sigma} / I_{\pi }$ ratio for the (0, 0, 2$\tau$) reflection hints to 
a possible symmetry break with loss of the 3-fold axis, however $ab$ $initio$ calculation would be needed 
to obtain quantitative informations. 
%
%
\begin{acknowledgments}
We would like to thank S. W. Lovesey and J. P. Hill for stimulating discussion. 
This work has been supported by the Swiss National Science Foundation, NCCR MaNEP.
\end{acknowledgments}
%
\appendix*
\section{Quadrupolar structure factor}
In analogy with Ref.~\onlinecite{ScagnoliPRB79} we obtain expression for $\Psi^K_Q$, 
written in the coordinate space (x,y,z), as a sum of quantities that are even ($A^K_Q$) and odd ($B^K_Q$) 
functions of the projection $Q$ with $-K \leq Q \leq  K$.\\
We give expression analog to Eq.~(B5) of Ref.~\onlinecite{ScagnoliPRB79} for a generic (0,0,$\ell$) reflection: 
\begin{equation}
A_{0}^{0}=\Psi_{0}^{0}
\end{equation}

\begin{eqnarray}
A_{0}^{1}&=&\frac{1}{\sqrt {2}} \left( \Psi_{-1}^{1}-\Psi_{1}^{1} \right) \\\nonumber
A_{1}^{1}&=&\frac{1}{2}\,(\Psi_{-1}^{1}+\,\Psi_{1}^{1}) \\\nonumber
B_{1}^{1}&=&\frac{1}{\sqrt {2}}\Psi_{0}^{1}\nonumber
\end{eqnarray}
\begin{eqnarray}{\label{eq:eqquadru}}
A_{0}^{2}&=&\frac{\sqrt{6}}{4}\, \left( \Psi_{-2}^{2}+\Psi_{2}^{2} \right) -\frac{1}{2}\,\Psi_{0}^{2}\\\nonumber
A_{1}^{2}&=&\frac{1}{2}\,(\Psi_{-2}^{2} - \Psi_{2}^{2}) \\\nonumber
B_{1}^{2}&=&\frac{1}{2}\,(\Psi_{-1}^{2} - \Psi_{1}^{2})\\\nonumber
A_{2}^{2}&=&\frac{1}{4}\,(\Psi_{-2}^{2} + \Psi_{2}^{2})+\frac{\sqrt {6}}{4}\,\Psi_{0}^{2}\\\nonumber
B_{2}^{2}&=&\frac{1}{2}\,(\Psi_{-1}^{2} + \Psi_{1}^{2}) \nonumber
\end{eqnarray}
Limiting ourselves to the quadrupolar contribution ($K$=2) and taking advantage of the 
structure factor $\Psi^2_Q (0,0,2\tau) = (-1)^{2\tau} \langle T^2_Q \rangle [ 1+2 \cos(Q \alpha)]$ we have only $\Psi_{0}^{2}$ different from zero.\\ 
Expressions in Eq.~(\ref{eq:eqquadru}) therefore simplify leading to e.g. $B_{Q}^{2}$ =0 and $A_{2}^{2} \propto A_{0}^{2}$. 
Substituting  Eq.~(\ref{eq:eqquadru}) in Eq.~(C1-C3) of Ref.~\onlinecite{ScagnoliPRB79} one obtains the 
expression quoted in Eq.~(\ref{eq:f2tau}). 
%
%
%

\begin{thebibliography}{56}
\bibitem [{vec()}]{vectordrop}%
  \BibitemOpen
  \href@noop {} {}\bibinfo {note} {We drop the bold style for $\bm{\epsilon}$
  ($\bm{\epsilon}^{\prime}$) to $\epsilon$ (${\epsilon}^{\prime}$) for the sake
  of simplicity.}\BibitemShut {Stop}%
\end{thebibliography}
%

\end{document}